\pdfoutput=1
\documentclass{article}
\usepackage{spconf,amsmath,graphicx}
\usepackage{cite}
\usepackage{setspace}
\usepackage[table]{xcolor}
\usepackage{bibspacing}
\usepackage{url}

\title{A Computationally Efficient Semi-Blind Source Separation Approach for Nonlinear Echo Cancellation Based on an Element-Wise Iterative Source Steering}
\name{Kunxing Lu$^{\star}$, Xianrui Wang$^{\star \dagger}$, Tetsuya Ueda$^{\star}$, Shoji Makino$^{\star}$, and Jingdong Chen$^{\dagger}$\thanks{This work was supported by JSPS KAKENHI Grant Number 23H03423.\\The audio samples in this research are available at {\protect\url{https://github.com/kunxinglu/audio_samples_ICASSP2024.}}}}
\address{$^{\star}$ Waseda University, Japan\\
$^{\dagger}$ Northwestern Polytechnical University, Xi'an, China}
\begin{document}
\ninept

\maketitle
\begin{spacing}{0.98}
\begin{abstract}
While the semi-blind source separation-based acoustic echo cancellation~(SBSS-AEC) has received much research attention due to its promising performance during double-talk compared to the traditional adaptive algorithms, it suffers from system latency and nonlinear distortions.
To circumvent these drawbacks, the recently developed ideas on convolutive transfer function~(CTF) approximation and nonlinear expansion have been used in the iterative projection~(IP)-based semi-blind source separation~(SBSS) algorithm.
However, because of the introduction of CTF approximation and nonlinear expansion, this algorithm becomes computationally very expensive, which makes it difficult to implement in embedded systems.
Thus, we attempt in this paper to improve this IP-based algorithm, thereby developing an element-wise iterative source steering~(EISS) algorithm.
In comparison with the IP-based SBSS algorithm, the proposed algorithm is computationally much more efficient, especially when the nonlinear expansion order is high and the length of the CTF filter is long.
Meanwhile, its AEC performance is as good as that of IP-based SBSS.
\end{abstract}
\begin{keywords}
Semi-blind source separation, acoustic
echo cancellation, convolutive transfer function approximation, nonlinear expansion, element-wise source steering
\end{keywords}
\section{Introduction}
\label{sec:intro}
In telecommunications or teleconferencing, acoustic echo, which is formed by the coupling between loudspeakers and microphones, is detrimental to full-duplex communication.
One widely used way to eliminate the detrimental echo effects is through acoustic echo cancellation in which adaptive filters, such as the normalized least mean square~(NLMS), recursive least mean square~(RLS) and Kalman filters~\cite{jacob2001,kalman}, are used to identify the acoustic impulse response~(AIR)~\cite{wang2020time}. 
While they have been widely used, adaptive filters generally suffer from great performance degradation or even divergence during double-talk in which both the far-end and near-end speech are present~\cite{gansler2000double,buchner2006dt}. 

One way to improve acoustic echo cancellation~(AEC) performance during double-talk is through adopting the principle used in blind source separation~(BSS)~\cite{comon1994ica,kim2006iva,ono2011auxiva,makino2018} to reformulate the AEC problem as one of semi-blind source separation~(SBSS)~\cite{sbsspropose1,sbsspropose2}.
Several algorithms have been derived based on this SBSS framework, which have demonstrated promising performance~\cite{gunther2011semi,koldovsky2014semi,gunther2015semi}.
However, those algorithms still suffer from a number of drawbacks. First, they are computationally very expensive, which makes them difficult to implement in embedded systems.
A viable approach to reduce complexity is through using the so-called multiplicative transfer function model~(MTF)~\cite{avargel2007mtf} and performing echo cancellation in the short-time Fourier transform~(STFT) domain~\cite{wada2008sbss,nesta2010sbss}.
But this method increases the system latency as MTF requires the length of the analysis window to be longer than the effective part of AIR.
Recently, the so-called convolutive transfer function~(CTF)~\cite{ctf1,talmon2009ctf,wang2023spatially} model was applied to SBSS-AEC to achieve different compromise between the system latency and computational complexity~\cite{wang2021,sbsscheng}.
Second, the performance of SBSS-AEC suffers from significant degradation in the presence of loudspeaker nonlinearity,
which happens often in small devices~\cite{niemisto2003performance,mossi2010assessment}.
One way to deal with this issue is through nonlinear AEC in which nonlinear expansion~\cite{malik2011fourier,Malik2012statespace,park2019} is used to model the loudspeaker nonlinearity~\cite{cheng2021semi}.
In~\cite{sbsscheng}, researchers proposed a framework which combines the CTF model and nonlinear expansion. An iterative projection~(IP)~\cite{ono2011auxiva} method is carried out to solve the corresponding optimization problem.

However, the IP-based algorithm still faces the challenge of computational complexity, as it requires to calculate the inverse of an auxiliary matrix. To circumvent this problem, we attempt in this work to reduce the complexity, thereby developing an element-wise source steering~(EISS) algorithm, which is an extension of the work in~\cite{robin2020iss,nakashima2022inverse,nakashima2023fast}. Since no matrix inverse is required, the developed algorithm is computationally much more efficient and its complexity is one order of magnitude lower than that of the IP-based algorithm, yet its performance is as good as that of the IP-based algorithm.

\section{Signal Model and Problem Formulation}
\label{sec:signal_model}

Consider the full-duplex speech communication scenario where a microphone is used to pick up the sound signal from the near-end speaker and a loudspeaker is used to playback the signal from the far-end. The microphone output signal at time instant $t$, which is denoted as $y(t)$, can be written as
\begin{align}
\nonumber
y(t) &= v(t) + s(t),\\
 &= a(t)*f\left[x(t)\right]+ s(t),
\label{Time_Echo}
\end{align}
where $s(t)$ denotes the near-end speech signal, $v(t)=a(t)*f\left[x(t)\right]$ is the nonlinear acoustic echo, $a(t)$ denotes the acoustic impulse response from the loudspeaker to the microphone, $*$ represents the linear convolution, $f(\cdot)$ stands for the response of the loudspeaker, which includes both the linear and nonlinear effects, and $x(t)$ is the far-end signal, respectively. The problem of acoustic echo cancellation is to mitigate or eliminate the echo signal, i.e., $v(t)$, while preserving the near-end signal $s(t)$.

While they have been widely used in practical systems, adaptive filtering algorithms often suffer from great performance degradation in the presence of loundspeaker nonlinearity. One way to deal with loundspeaker nonlinear distortion is to  approximate $f\left[x(t)\right]$ through a $P$th-order basis-generic expansion~\cite{Malik2012statespace}, i.e., 
\begin{equation}
f\left[x(t)\right]= \sum_{p=0}^{P-1}c_p\phi_{p}\left[x\left(t\right)\right],\quad p=0,1,\ldots, P-1,
\label{Non_Expan}
\end{equation}
where $\phi_{p}\left(\cdot\right)$ is the $p$-th order basis function and $c_p$ is the corresponding coefficient. For real-valued signal, the following expansion can be used,
\begin{equation}
\phi_p\left[x\left(t\right)\right]=x^{2p+1}(t).
\end{equation}
Substituting \eqref{Non_Expan} into \eqref{Time_Echo} gives
\begin{align}
\nonumber
y(t) &= \sum_{p=0}^{P-1}c_p a(t) * \phi_p\left[x\left(t\right)\right] + s(t),\\
&= \sum_{p=0}^{P-1}a^{\prime}_p(t) *\phi_p\left[x(t)\right] + s(t),
\label{Non_Line}
\end{align}
where $a^{\prime}_p(t)= c_p a(t)$ denote the echo path of the $p$-th order expanded signal $\phi_p\left[x(t)\right]$.

To reduce the computational complexity, the MTF model is introduced, which requires the analysis window to be longer than the effective part of AIR, leading to larger system latency. To achieve proper compromise between the computational complexity and system latency, the CTF approximation is subsequently adopted. The signal model in (\ref{Non_Line}) is then written in the STFT domain as
\begin{equation}
Y_{i,j} = \sum _{p=0}^{P-1} \sum _{l=0}^{L-1} A ^{\prime}_{p,i,l} X_{\phi,p,i,j-l} + S_{i,j}, 
\label{Freq_mix}
\end{equation}
where $i$ and $j$ denote the frequency and time-frame indexes, $L$ is the length of the CTF filter, and $Y_{i,j}$, $A^{\prime}_{p,i,j}$, $X_{\phi, p,i,j}$, $S_{i,j}$ denote, respectively, the STFTs of $y(t)$, $a^{\prime}_p(t)$, $\phi_p\left[x(t)\right]$ and $s(t)$. Putting \eqref{Freq_mix} into a vector/matrix form gives
\begin{equation}
\tilde{\mathbf y}_{i,j} = \tilde{\mathbf{H}}_{i,j} \tilde{\mathbf{s}}_{i,j},
\label{matrix_mix}
\end{equation}
where
\begin{equation}
\tilde{\mathbf{y}}_{i,j} = \left[Y_{i,j} \quad \mathbf x^T_{0,i,j} \quad \cdots \quad \mathbf x^T_{P-1,i,j}\right]^T,\\
\end{equation}

\begin{equation}
Y_{i,j} = \sum _{p=0}^{P-1} \mathbf a^T_{p,i,j} \mathbf x_{p,i,j} + S_{i,j},
\label{vector_mix}
\end{equation}

\begin{equation}
\mathbf a_{p,i,j} = [A^\prime_{p,i,0}\quad \cdots \quad A^\prime_{p,i,L-1}]^T,\\
\end{equation}

\begin{equation}
\mathbf x_{p,i,j} =[X_{\phi,p,i,j}\quad \cdots \quad X_{\phi,p,i,j-L+1}]^T,\\
\end{equation}

\begin{equation}
\tilde{\mathbf{s}}_{i,j} = \left[S_{i,j} \quad \mathbf x^T_{0,i,j} \quad \cdots \quad \mathbf x^T_{P-1,i,j}\right]^T,\\
\end{equation}


\begin{equation}
\tilde{\mathbf{H}}_{i, j} = \begin{bmatrix}
1 & \mathbf{a}^T_{i, j} \\
\\
\mathbf{0}_{PL \times 1} & \mathbf{I}_{PL}
\end{bmatrix},\\
\end{equation}

\begin{equation}
\mathbf{a}_{i, j} =\left[\mathbf a^T_{0,i,j}\quad \mathbf a^T_{1,i,j}\quad \ldots\quad \mathbf a^T_{P-1,i,j}\right]^T,\\
\end{equation}
the superscript $T$ denotes the transpose operation, and $\mathbf I_{PL}$ denotes the identity matrix of size $PL \times PL$. 
Note that the $\tilde{\mathbf{H}}_{i, j}$ matrix is of size $(PL+1) \times (PL+1)$, which is called the mixing matrix in the literature of BSS, $\mathbf{0}_{PL \times 1}$ is a zero vector of length $PL$. 

Following the notation in BSS and echo cancellation, we now define the demixing matrix $\tilde{\mathbf{W}}_{i,j}$ as 
\begin{equation}
\tilde{\mathbf{W}}_{i,j}=\left[\begin{array}{cc}
1 & \mathbf{b}^T_{i,j} \\
\\
\mathbf{0}_{PL \times 1} & \mathbf{I}_{PL}
\end{array}\right],
\end{equation}
where $\mathbf{b}_{i,j}$ is a column vector with $PL$ parameters to be estimated. The near-end signal extraction filter can then be expressed as $\tilde{\mathbf{w}}_{i,j}^H=[1 \quad \mathbf b_{i,j}^T]$. Applying the near-end signal extraction filter to the input signal gives the near-end signal, i.e., 
\begin{equation}
\hat S_{i,j} = \tilde{\mathbf{w}}^H_{i,j}\tilde{\mathbf y}_{i,j}.
\end{equation}
Given the aforementioned signal model and problem formulation, the objective of nonlinear SBSS-AEC is to estimate $\tilde{\mathbf{w}}_{i,j}$ by exploiting independence between the near-end and the reference signals.

\section{Nonlinear SBSS-AEC Algorithms}
\label{sec:SBSS_algorithm}
\subsection{Probabilistic Model}
We consider to model the source signal with a generalized Gaussian distribution, i.e.,
\begin{equation}
p\left(\mathbf{s}_j\right) \propto \exp \left[-\left(\frac{\left\|\mathbf{s}_j\right\|_2}{\gamma}\right)^\beta\right],
\end{equation}
where
\begin{equation}
\mathbf{s}_j = \left[S_{1,j}\quad S_{2,j}\quad \ldots \quad S_{I,j}\right]^T,
\end{equation}
$\left\|\cdot \right\|_2$ stands for $\ell_2$ norm. $\gamma$ and $\beta$ are the scale and shape parameters, respectively. Since the reference signal is accessible, the negative log-likelihood function can then be calculated as
\begin{equation}
\begin{aligned}
\begin{split}
\mathcal{L}_j = &-\frac{1}{\sum_{j^{\prime}=1}^j\alpha^{j-j^{\prime}}}\sum_{j^{\prime}=1}^j \alpha^{j-j^{\prime}}\log p\left(\mathbf{s}_{j^{\prime}}\right) \\ &\qquad\qquad\qquad\qquad- 2\sum_{i=1}^{I}\log\vert\det\tilde{\mathbf{W}}_{i,j}\vert,
\label{loss_function}
\end{split}
\end{aligned}
\end{equation}
where $\alpha \in (0, 1)$ is a forgetting factor. By using the well known majorization-minimization~(MM) method~\cite{lange2016mm}, the following auxiliary function can be obtained
\begin{equation}
\mathcal{L}^+_j = \sum _{i=1}^I \tilde{\mathbf{w}}_{i,j}^H \mathbf V_{i,j} \tilde{\mathbf{w}}_{i,j} -2\sum _{i=1}^I\log \left|\det\tilde{\mathbf W}_{i,j}\right|.
\label{aux_function}
\end{equation}
To track time-varying signals and acoustic environments, the recursive estimation of $\mathbf V_{i,j}$ is generally used, i.e.,
\begin{equation}
\mathbf{V}_{i,j} = \alpha \mathbf V_{i,j-1} + (1-\alpha)\varphi(r_{j}) \tilde{\mathbf{y}}_{i,j}\tilde{\mathbf{y}}_{i,j}^H,
\end{equation}
and
\begin{eqnarray}
&\varphi(r_j) = r_j^{\beta-2},\\
&r_j = \sqrt{\sum_{i=1}^{I}\vert\tilde{\mathbf{w}}^H_{i,j-1}\tilde{\mathbf{y}}_{i,j}\vert^2}.
\end{eqnarray}
Note that the norm of the near-end signal extraction filter $\tilde{\mathbf{w}}_{i,j}$ does not affect the independence criterion. Therefore, a two-stage estimation strategy can be used in which the extraction filter is updated in the first stage using the maximum likelihood criterion and the updated filter is then normalized in the second stage such that its first element is equal to 1. Since the extraction filters at different frequency bins are estimated independently, we shall omit the frequency index $i$ in the rest parts of this paper without introducing any confusion.
\subsection{Conventional Iterative Projection Based Method}
The IP method can be derived by identifying the Wirtinger derivative of \eqref{aux_function} with respect to $\tilde{\mathbf{w}}_{i,j}$ and then forcing the result equal to $0$. The update rules are shown as follows:
\begin{align}
&\tilde{\mathbf w}_{j} \leftarrow (\tilde{\mathbf{W}}_{j-1}\mathbf{V}_{j})^{-1} \mathbf{e}_1 = \mathbf{V}_j^{-1} \mathbf{e}_1, \\
&\tilde{\mathbf w}_{j} \leftarrow \frac {\tilde{\mathbf w}_{j}}{w_{1,j}},
\end{align}
where $\mathbf{e}_1$ is the first column of $\mathbf{I}_{PL+1}$ and $w_{1,j}$ is the first element of $\tilde{\mathbf w}_{j}$.

\subsection{Proposed Element-wise Iterative Source Steering Method}
\begin{figure}[t!]
 \centering
 \includegraphics[width=75mm]{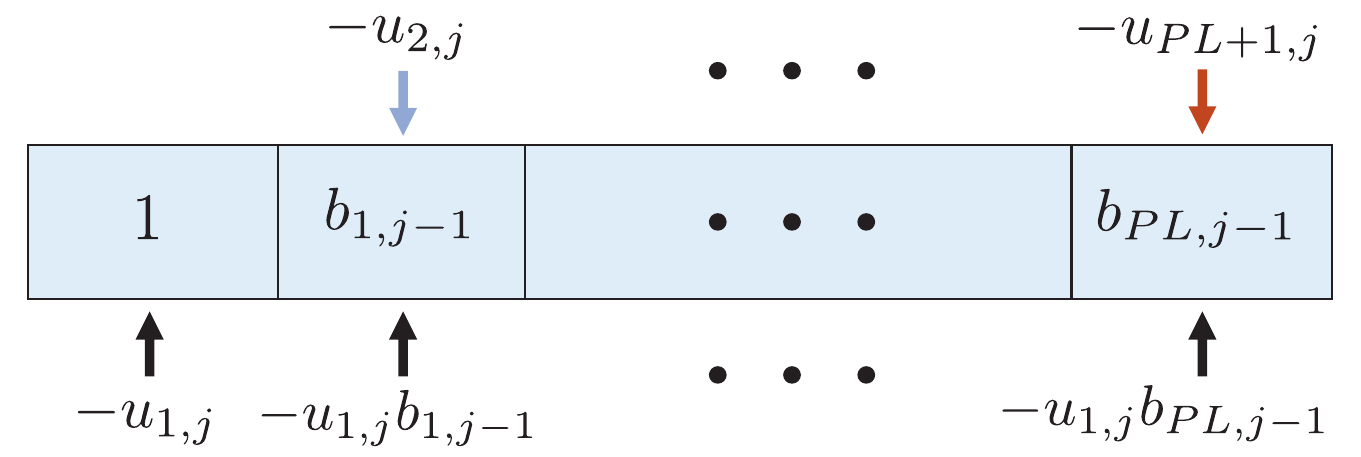}
 \caption{Illustration of EISS update rule.}
 \label{update}
\end{figure}

Implementation of the IP method requires to compute the inverse of the $\mathbf{V}_{j}$ matrix every frame, which makes the algorithm computationally very expensive. To reduce the complexity, we propose to update the near-end signal extraction filter with the EISS method in the first stage~\cite{robin2020iss,nakashima2022inverse,nakashima2023fast}, i.e., 
\begin{equation}
\begin{cases}
w_{1,j} \leftarrow w_{1,j-1} - u_{1,j}, & \text{if } k = 1 \\
w_{k,j} \leftarrow w_{k,j-1}- u_{1,j} w_{k,j-1}-u_{k,j}, & \text{if }k\ne1
\label{update_rule}
\end{cases}
\end{equation}
where $w_{k,j}, ~k=1,2,\ldots, PL+1$, is $k$-th element of $\tilde{\mathbf{w}}_{j}$ and $u_{k,j}$ is a parameter to estimate. All the $u_{k,j}$'s need to be estimated sequentially. In other words, the algorithm first computes $u_{1,j}$ to update all the elements related to $w_{j-1}$; it then computes $u_{2,j}$ to update $b_{1,j-1}$; it subsequently computes $u_{3,j}$ to update $b_{2,j-1}$, and so forth. The EISS update rules are illustrated in Fig.~\ref{update}.

Substituting \eqref{update_rule} into the auxiliary function, we obtain
\begin{equation}
\begin{split}
\mathcal{L}_{j}^+(u_{k,j}) =& -2\log|1 - u_{1,j}|+ \mathbf d_{j}^H \mathbf V_{j} \mathbf d_{j},
\end{split}
\end{equation}
where
\begin{equation}
\begin{split}
\begin{aligned}
\mathbf d_{j}^H = &\tilde{\mathbf{w}}^H_{j-1} - \left[u_{1,j}\quad u_{1,j}b_{1,j-1}+u_{2,j}\quad \right.\\&\left. \qquad\qquad\qquad \cdots  u_{1,j}b_{PL,j-1}+u_{PL+1,j} \right].
\end{aligned}
\end{split}
\end{equation}
Identifying the Wirtinger derivative of $\mathcal{L}_{j}^+(u_{k,j})$ with respect to $u_{k,j}^*$ and forcing the result to be $0$, one can obtain the following solution
\begin{equation}
u_{k,j}= \begin{cases}\begin{aligned}
&\frac {\tilde{\mathbf{w}}_{j-1}^H \mathbf v_{k,j}}{V_{j}(k,k)},&\quad k \neq 1 \\
&1-(\tilde{\mathbf{w}}_{j-1}^H \mathbf V_{j} \tilde{\mathbf{w}}_{j-1})^{-\frac {1}{2}},&\quad k=1
\end{aligned}
\end{cases},
\end{equation}
where $\mathbf v_k$ denotes the $k$-th column of $\mathbf V_{j}$ and ${V_{j}(k,k)}$ stands for the $(k, k)$-th element of $\mathbf{V}_{j}$.

\section{Complexity analysis}
\label{sec:complexity_analysis}

In this section, we present the complexity of the IP and EISS algorithms in terms of the number of multiplications/divisions needed per frequency bin and time frame. For the IP algorithm, the complexity is dominated by computing the inverse of the covariance matrix $\mathbf{V}_j$, which has a complexity proportional to $\mathcal{O}(PL+1)^3$. The complexity for all the other computations is of $\mathcal{O}(PL+1)$. For regular setup, $\mathcal{O}(PL+1)$ is much smaller than $\mathcal{O}(PL+1)^3$. Therefore, the complexity of the IP method is
\begin{equation}
\mathcal{C}_\mathrm{IP} \propto \mathcal{O}\left[(PL)^3\right].
\end{equation}
The EISS algorithm needs to estimate $PL+1$ coefficients. Computation of $u_{k,j}, k\neq1$ has a complexity of $\mathcal{O}\left[PL(PL+1)\right]$. The complexity for computing $u_{1,j}$ is $\mathcal{O}\left[(PL+1)^2\right]$ and it is $O(PL+1)$
for all the other operations. Similarly, if we only consider the operations that dominate the complexity, the overall complexity of EISS method is
\begin{equation}
\mathcal{C}_\mathrm{EISS} \propto \mathcal{O}\left[(PL)^2\right],
\end{equation}
which is one order lower than that of the IP method.

\section{Simulations and Experiments}
\label{sec:simulations}
\subsection{Experimental Setup}
In this section, the proposed EISS-based algorithm is compared with the IP-based algorithm and the single-microphone form of the state-space model~(SSM)-based nonlinear acoustic echo cancellation algorithm proposed in \cite{park2019}.
We evaluated the proposed AEC algorithms with the help of objective measures to quantify the performance in terms of echo reduction and speech distortion. For the single-talk case, echo return loss enhancement~(ERLE)~\cite{Malik2012statespace} is used as the performance metric, and for double-talk,  true ERLE~(tERLE)~\cite{nesta2010sbss} is used. Besides, perceptual evaluation of speech quality~(PESQ)~\cite{pesq}, and short time objective intelligibility~(STOI)~\cite{stoi} are also used for performance evaluation. The sampling rate for all the signals in this work is 16~kHz.

To assess the efficiency of our algorithm, we also conducted a comparative analysis of the runtime performance between our algorithm and an IP-based algorithm.
We executed 100 signals, each lasting 10 seconds, on a laptop equipped with an i7-10750H CPU and computed the average runtime of each signal as the final test result.
The average runtime is showned in Fig.~\ref{runtime}.
For the short-time analysis, the frame length is 1024-point long with an overlap factor of $75\%$. The Hanning window is applied and the windowed signal is then transformed into the STFT domain with a 1024-point fast Fourier transform~(FFT).
To balance the computational complexity and performance, the nonlinear expansion order $P$ is set to 3 and the length of CTF filter $L$ is set to 5. The forgetting factor $\alpha$ is set to 0.992. The shape parameter $\beta$ is set to 0.4. In all experiments, the demixing matrix $\tilde{\mathbf{W}}$ is initialized as an identity matrix $\mathbf{I}$ and the auxiliary matrix $\mathbf V$ is initialized as $10^{-3} \times \mathbf I$.

\subsection{AEC Performance for Hard Clipping Mapping}

\begin{figure}[htbp]
\centering
\includegraphics[width=85mm]{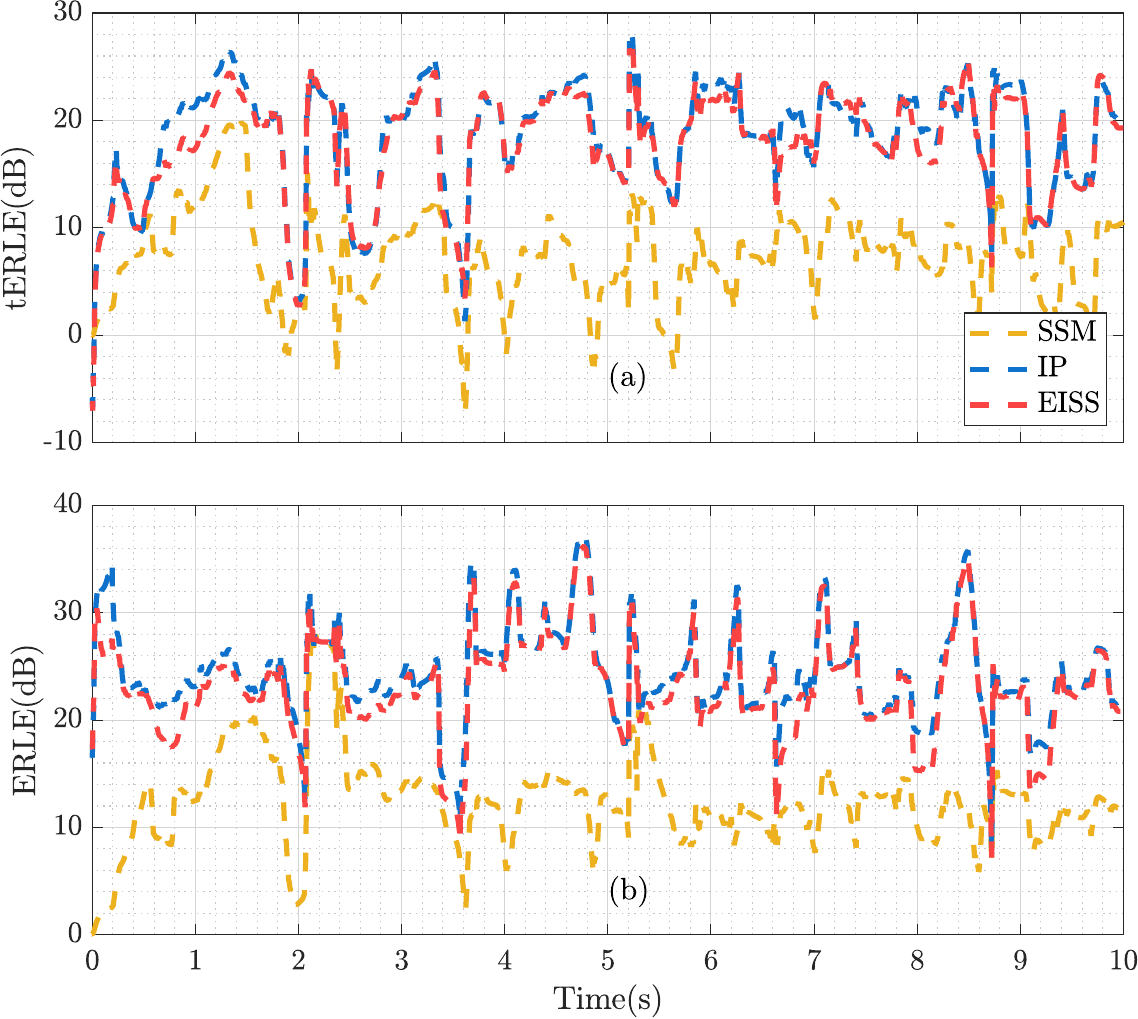}
\caption{The performance of SSM, IP and EISS: (a).~tERLE in double-talk situation, (b).~ERLE in single-talk situation.}
\label{ERLE}
\end{figure}

In this experiment, we validate the ability of EISS to handle nonlinear distortion in both single-talk and double-talk scenarios.
We use the same data as in~\cite{sbsscheng}. The hard clipping function~\cite{Malik2012statespace} is used to simulate the loudspeaker nonlinearity, in which the clipping threshold is set to $0.2max|x(t)|$. The reverberation time $T_{60}$ is approximately 300~ms. The signal-to-echo ratio~(SER) for double-talk is 0~dB. Figure~\ref{ERLE}~(a) and~(b) show the performance of the IP and EISS methods in the double-talk and single-talk situations, respectively. One can see that the ERLE and tERLE of EISS and IP algorithms are almost the same, but when compared to SSM, both of them have significantly higher values. Therefore, the proposed algorithm demonstrates superior AEC performance compared to the conventional SSM algorithm.

\subsection{Overall Performance on the AEC Challenge Dataset}

In this experiment, a total of 30 signals are arbitrarily taken from the AEC challenge synthetic dataset~\cite{aecchallenge}. The nonlinear far-end signals are generated by clipping the maximum amplitude or by applying the sigmoidal function~\cite{lee2015dnn} and learned distortion functions to far-end signal. The SER of these signals ranges from $-10$~dB to $10$~dB and the $T_{60}$ ranges from 200~ms to 1200~ms.
\begin{table}[htbp]
\vspace{-5pt}
\caption{Performance of SSM, IP and EISS.}
\vskip 4pt
\label{average value}
\centering
\setlength{\tabcolsep}{5mm}{
\renewcommand{\arraystretch}{1.1}
\begin{tabular}{cccc}
\hline
Algorithm& PESQ&STOI&tERLE \\
\hline
SSM&1.57& 0.87& 8.77\\
IP& 1.89& 0.93&12.89\\
\rowcolor{gray!40}
EISS& 1.9& 0.94&12.63\\
\hline
\end{tabular}}
\end{table}
Table~\ref{average value} lists the overall performance of the three algorithms in terms of PESQ, STOI, and tERLE. It can be seen that the proposed algorithm significantly outperforms traditional nonlinear AEC algorithms across various noise and reverberation environments and exhibits similar performance to the IP-based algorithm.
\subsection{Runtime Comparison}

In the last set of experiments, we compare the runtime of the IP and EISS method with the same setup as described previously. The time measured here includes the time to compute and update the covariance matrix and auxiliary variables. In this experiment, the nonlinear expansion order $P$ is set to 3 and 4, respectively. The CTF filter length, i.e., $L$, varies from $2$ to $12$.
\begin{figure}[htbp]
\centering
\includegraphics[width=85mm]{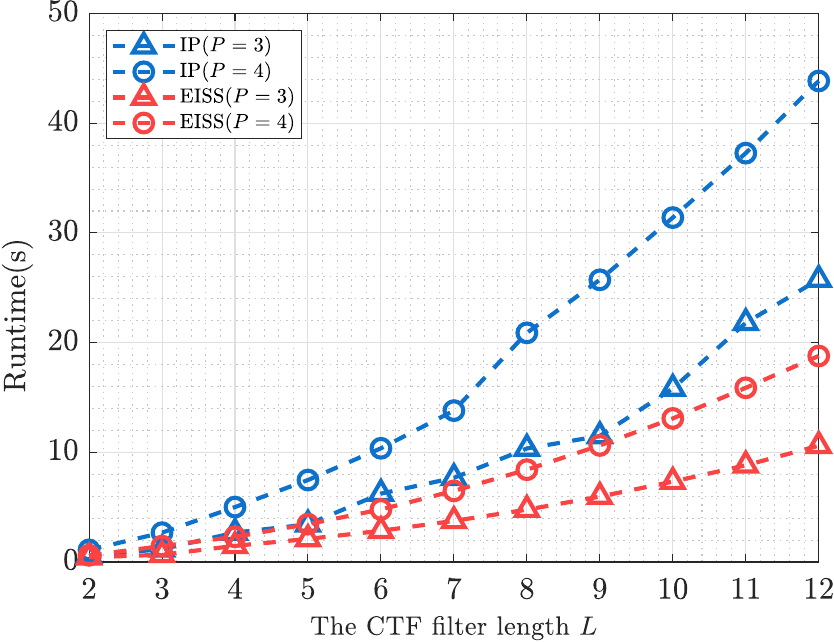}
\caption{The runtime comparison between IP and EISS.}
\label{runtime}
\end{figure}
As shown in Fig.~\ref{runtime}, the EISS method is computationally much more efficient than the IP method and the difference becomes much more dramatic as values of $P$ and $L$ increases.

\section{CONCLUSION}

This paper investigated the problem of AEC in the presence of doubletalk and loudspeaker nonlinearity within reverberant environment. Following the framework of SBSS, we developed an element-wise source steering algorithm, which combines the CTF model and nonlinear expansion into the SBSS framework. Unlike the conventional IP-based SBSS method, which requires to compute the inverse of the auxiliary matrix, the proposed algorithm applies an element-wise update strategy, in which no matrix inverse is involved, and as a result, its computational complexity is one order of magnitude lower than that of the IP-based algorithm. Moreover, experiments showed that this efficient algorithm is able to achieve similar performance to that of the IP-based algorithm.
\end{spacing}

\vfill\pagebreak



\begin{thebibliography}{10}
\providecommand{\url}[1]{#1}
\csname url@samestyle\endcsname
\providecommand{\newblock}{\relax}
\providecommand{\bibinfo}[2]{#2}
\providecommand{\BIBentrySTDinterwordspacing}{\spaceskip=0pt\relax}
\providecommand{\BIBentryALTinterwordstretchfactor}{4}
\providecommand{\BIBentryALTinterwordspacing}{\spaceskip=\fontdimen2\font plus
\BIBentryALTinterwordstretchfactor\fontdimen3\font minus
  \fontdimen4\font\relax}
\providecommand{\BIBforeignlanguage}[2]{{%
\expandafter\ifx\csname l@#1\endcsname\relax
\typeout{** WARNING: IEEEtran.bst: No hyphenation pattern has been}%
\typeout{** loaded for the language `#1'. Using the pattern for}%
\typeout{** the default language instead.}%
\else
\language=\csname l@#1\endcsname
\fi
#2}}
\providecommand{\BIBdecl}{\relax}
\BIBdecl

\bibitem{jacob2001}
J.~Benesty, T.~G{\"a}nsler, D.~R. Morgan, M.~M. Sondhi, S.~L. Gay
  \emph{et~al.}, \emph{Advances in network and acoustic echo
  cancellation}.\hskip 1em plus 0.5em minus 0.4em\relax Springer, 2001.

\bibitem{kalman}
G.~Enzner and P.~Vary, ``Frequency-domain adaptive {K}alman filter for acoustic
  echo control in hands-free telephones,'' \emph{Signal Process.}, vol.~86,
  no.~6, pp. 1140--1156, 2006.

\bibitem{wang2020time}
X.~Wang, G.~Huang, J.~Benesty, J.~Chen, and I.~Cohen, ``Time difference of
  arrival estimation based on a {K}ronecker product decomposition,'' \emph{IEEE
  Signal Process. Lett.}, vol.~28, pp. 51--55, 2020.

\bibitem{gansler2000double}
T.~Gansler, S.~L. Gay, M.~M. Sondhi, and J.~Benesty, ``Double-talk robust fast
  converging algorithms for network echo cancellation,'' \emph{IEEE Trans.
  Speech, Audio Process.}, vol.~8, no.~6, pp. 656--663, 2000.

\bibitem{buchner2006dt}
H.~Buchner, J.~Benesty, T.~Gansler, and W.~Kellermann, ``Robust extended
  multidelay filter and double-talk detector for acoustic echo cancellation,''
  \emph{IEEE Trans. Audio, Speech, Lang. Process.}, vol.~14, no.~5, pp.
  1633--1644, 2006.

\bibitem{comon1994ica}
P.~Comon, ``Independent component analysis, a new concept?'' \emph{Signal
  Process.}, vol.~36, no.~3, pp. 287--314, 1994.

\bibitem{kim2006iva}
T.~Kim, H.~T. Attias, S.-Y. Lee, and T.-W. Lee, ``Blind source separation
  exploiting higher-order frequency dependencies,'' \emph{IEEE Trans. Audio,
  Speech, Lang. Process.}, vol.~15, no.~1, pp. 70--79, 2006.

\bibitem{ono2011auxiva}
N.~Ono, ``Stable and fast update rules for independent vector analysis based on
  auxiliary function technique,'' in \emph{Proc. WASPAA}, 2011, pp. 189--192.

\bibitem{makino2018}
S.~Makino, \emph{Audio source separation}.\hskip 1em plus 0.5em minus
  0.4em\relax Springer, 2018.

\bibitem{sbsspropose1}
M.~Joho, H.~Mathis, and G.~S. Moschytz, ``Combined blind/nonblind source
  separation based on the natural gradient,'' \emph{IEEE Signal Process.
  Lett.}, vol.~8, no.~8, pp. 236--238, 2001.

\bibitem{sbsspropose2}
S.~Miyabe, T.~Takatani, H.~Saruwatari, K.~Shikano, and Y.~Tatekura,
  ``Barge-in-and noise-free spoken dialogue interface based on sound field
  control and semi-blind source separation,'' in \emph{Proc. EUSIPCO}, 2007,
  pp. 232--236.

\bibitem{gunther2011semi}
J.~Gunther, ``Learning echo paths during continuous double-talk using
  semi-blind source separation,'' \emph{IEEE Trans. Audio, Speech, Lang.
  Process.}, vol.~20, no.~2, pp. 646--660, 2011.

\bibitem{koldovsky2014semi}
Z.~Koldovsk{\`y}, J.~M{\'a}lek, M.~M{\"u}ller, and P.~Tichavskj{\`y}, ``On
  semi-blind estimation of echo paths during double-talk based on
  nonstationarity,'' in \emph{Proc. IWAENC}, 2014, pp. 198--202.

\bibitem{gunther2015semi}
J.~Gunther and T.~Moon, ``Blind acoustic echo cancellation without double-talk
  detection,'' in \emph{Proc. WASPAA}, 2015, pp. 1--5.

\bibitem{avargel2007mtf}
Y.~Avargel and I.~Cohen, ``On multiplicative transfer function approximation in
  the short-time {F}ourier transform domain,'' \emph{IEEE Signal Process.
  Lett.}, vol.~14, no.~5, pp. 337--340, 2007.

\bibitem{wada2008sbss}
T.~S. Wada, S.~Miyabe, and B.-H.~F. Juang, ``Use of decorrelation procedure for
  source and echo suppression,'' in \emph{Proc. IWAENC}, 2008, pp. 1--5.

\bibitem{nesta2010sbss}
F.~Nesta, T.~S. Wada, and B.-H. Juang, ``Batch-online semi-blind source
  separation applied to multi-channel acoustic echo cancellation,'' \emph{IEEE
  Trans. Audio, Speech, Lang. Process.}, vol.~19, no.~3, pp. 583--599, 2010.

\bibitem{ctf1}
R.~Talmon, I.~Cohen, and S.~Gannot, ``Relative transfer function identification
  using convolutive transfer function approximation,'' \emph{IEEE Trans. Audio,
  Speech, Lang. Process.}, vol.~17, no.~4, pp. 546--555, 2009.

\bibitem{talmon2009ctf}
R.~Talmon, I.~Cohen, and S.~Gannot, ``Convolutive transfer function generalized sidelobe canceler,''
  \emph{IEEE Trans. Audio, Speech, Lang. Process.}, vol.~17, no.~7, pp.
  1420--1434, 2009.

\bibitem{wang2023spatially}
X.~Wang, A.~Brendel, G.~Huang, Y.~Yang, W.~Kellermann, and J.~Chen, ``Spatially
  informed independent vector analysis for source extraction based on the
  convolutive transfer function model,'' in \emph{Proc. IEEE ICASSP}, 2023, pp.
  1--5.

\bibitem{wang2021}
Z.~Wang, Y.~Na, Z.~Liu, B.~Tian, and Q.~Fu, ``Weighted recursive least square
  filter and neural network based residual echo suppression for the
  {AEC}-challenge,'' in \emph{Proc. IEEE ICASSP}, 2021, pp. 141--145.

\bibitem{sbsscheng}
G.~Cheng, L.~Liao, K.~Chen, Y.~Hu, C.~Zhu, and J.~Lu, ``Semi-blind source
  separation using convolutive transfer function for nonlinear acoustic echo
  cancellation,'' \emph{J. Acoust. Soc. Am.}, vol. 153, no.~1, pp. 88--95,
  2023.

\bibitem{niemisto2003performance}
R.~Niemist{\"o} and T.~M{\"a}kel{\"a}, ``On performance of linear adaptive
  filtering algorithms in acoustic echo control in presence of distorting
  loudspeakers,'' in \emph{Proc. IWAENC}, 2003, pp. 79--82.

\bibitem{mossi2010assessment}
M.~I. Mossi, N.~W. Evans, and C.~Beaugeant, ``An assessment of linear adaptive
  filter performance with nonlinear distortions,'' in \emph{Proc. IEEE ICASSP},
  2010, pp. 313--316.

\bibitem{malik2011fourier}
S.~Malik and G.~Enzner, ``{Fourier} expansion of {Hammerstein} models for
  nonlinear acoustic system identification,'' in \emph{Proc. IEEE ICASSP},
  2011, pp. 85--88.

\bibitem{Malik2012statespace}
------, ``State-space frequency-domain adaptive filtering for nonlinear
  acoustic echo cancellation,'' \emph{IEEE Trans. Audio, Speech, Lang.
  Process.}, vol.~20, no.~7, pp. 2065--2079, 2012.

\bibitem{park2019}
J.~Park and J.-H. Chang, ``State-space microphone array nonlinear acoustic echo
  cancellation using multi-microphone near-end speech covariance,''
  \emph{IEEE/ACM Trans. Audio, Speech, Lang. Process.}, vol.~27, no.~10, pp.
  1520--1534, 2019.

\bibitem{cheng2021semi}
G.~Cheng, L.~Liao, H.~Chen, and J.~Lu, ``Semi-blind source separation for
  nonlinear acoustic echo cancellation,'' \emph{IEEE Signal Process. Lett.},
  vol.~28, pp. 474--478, 2021.

\bibitem{robin2020iss}
R.~Scheibler and N.~Ono, ``Fast and stable blind source separation with rank-1
  updates,'' in \emph{Proc. IEEE ICASSP}, 2020, pp. 236--240.

\bibitem{nakashima2022inverse}
T.~Nakashima and N.~Ono, ``Inverse-free online independent vector analysis with
  flexible iterative source steering,'' in \emph{Proc. APSIPA}, 2022, pp.
  749--753.

\bibitem{nakashima2023fast}
T.~Nakashima, R.~Ikeshita, N.~Ono, S.~Araki, and T.~Nakatani, ``Fast online
  source steering algorithm for tracking single moving source using online
  independent vector analysis,'' in \emph{Proc. IEEE ICASSP}, 2023, pp. 1--5.

\bibitem{lange2016mm}
K.~Lange, \emph{MM optimization algorithms}.\hskip 1em plus 0.5em minus
  0.4em\relax SIAM, 2016.

\bibitem{pesq}
A.~W. Rix, J.~G. Beerends, M.~P. Hollier, and A.~P. Hekstra, ``Perceptual
  evaluation of speech quality ({PESQ})-a new method for speech quality
  assessment of telephone networks and codecs,'' in \emph{Proc. IEEE ICASSP},
  vol.~2, 2001, pp. 749--752.

\bibitem{stoi}
C.~H. Taal, R.~C. Hendriks, R.~Heusdens, and J.~Jensen, ``A short-time
  objective intelligibility measure for time-frequency weighted noisy speech,''
  in \emph{Proc. IEEE ICASSP}, 2010, pp. 4214--4217.

\bibitem{aecchallenge}
R.~Cutler, A.~Saabas, T.~Parnamaa, M.~Purin, H.~Gamper, S.~Braun,
  K.~S{\o}rensen, and R.~Aichner, ``{ICASSP} 2022 acoustic echo cancellation
  challenge,'' in \emph{Proc. IEEE ICASSP}, 2022, pp. 9107--9111.

\bibitem{lee2015dnn}
C.~M. Lee, J.~W. Shin, and N.~S. Kim, ``{DNN}-based residual echo
  suppression,'' in \emph{Proc. Interspeech}, 2015, pp. 1175--1179.

\end{thebibliography}
\end{document}